\documentclass[a4paper,11pt]{amsart}
\title{Towards an information-theoretically safe cryptographic protocol}
\newtheorem{theorem}{Theorem}
\newtheorem*{theorem*}{Theorem}
\newtheorem{conjecture}{Conjecture}
\newtheorem{question}{Question}
\newtheorem{definition}{Definition}
\author{Pedro Fortuny Ayuso}
\date{March 24, 2006}
\email{pfortuny@sdf-eu.org}
\begin{document}
\begin{abstract}
  We introduce what --if some kind of group action exists-- is a truly
  (information-theoretically) safe cryptographic communication system: a 
  protocol which provides \emph{zero} information to any
  passive adversary having full access to the channel.
\end{abstract}
\maketitle
\section{The false algorithm, simple version}\label{sec:false-algorithm}
Assume Alice wants to share a secret $s$, which we assume for
simplicity\footnote{This assumption might be relaxed, using an
  infinite set is for exposition reasons, see section \ref{sec:requirements}.} 
is a non-zero rational number $s=p/q\in {\mathbb Q}^{\star}$. For example, $s$ could be the key of a 
symmetric key protocol, a password
or even a complete message such as a pair of
coordinates in a map or a time.\par
Alice picks another random rational $t$ and calls $v=(s,t)$ to
the corresponding point in ${\mathbb Q}^{2}$.\par
She chooses a random transformation $A\in GL_{2}({\mathbb Q})$ in the
linear group of ${\mathbb Q}^{2}$ and computes $v_{1}=v\cdot A$. Alice sends $v_{1}$
to Bob.\par
Bob picks another random transformation $B\in GL_{2}({\mathbb Q})$ and
computes $v_{2}=v_{1}\cdot B$, and sends $v_{2}$ back to Alice. Notice
that $v_{1}$ gives no information to Bob or an eavesdropper (Eve) about $s$,
because $t$ is
random and $v_{1}$ can be \emph{any} point in ${\mathbb Q}^{2}$, 
depending on $t$ and $A$, which
are both unknown to both Bob and Eve. For a similar reason, the knowledge of $v_{1}$ 
and $v_{2}$ gives no
useful information about $B$.\par
Alice now computes $v_{3}=v_{2}\cdot A^{-1}$ and sends $v_{3}$ back
to Bob. Again, the knowledge of $v_{1},$
$v_{2}$ and $v_{3}$ is useless in order to retrieve the original $v$.\par
Finally, Bob computes $v_{4}=v_{3}\cdot B^{-1}$.\par
If only $v_{4}=v$...!
\section{The protocol ``would be'' safe}
Let us assume the above algorithm ends up with $v_{4}=v$ and let us
prove its safeness under this condition.
\begin{theorem}
  The above method of communication is information-theoretically safe,
  assuming $v$, $A$
  and $B$ (and their inverses, obviously) are kept secret. That is,
  the knowledge of the whole communication gives no information on the message.
\end{theorem}
\begin{proof}
  We only need to show that an eavesdropper which knows all the
  communication has no clue about what $s$ may be. In other words, it
  is enough to show that for any rational $s^{\prime}$, there exist
  another rational number $t^{\prime}$ and matrices $A'$, $B'$ such
  that the communication between Alice and Bob is the same
  (i.e. $v_{1}, v_{2}$ and $v_{3}$). But this
  is trivial.
\end{proof}
\textbf{Remark:\/} The algorithm described above obviously does not work because
$GL_{2}({\mathbb Q})$ is non-commutative (in general, the linear group is noncommutative for dimension greater than $1$).
\section{What is needed?}\label{sec:requirements}
A natural question comes to mind: what are the necessary conditions for a
group action on a set for the above algorithm to provide a valid
system? What we used above is:
\begin{enumerate}
\item A set $S$ (either finite or infinite) (the rational plane in the
  example).
\item An action $G\times S^{2}\rightarrow S^{2}$ of a \emph{commutative} group $G$
  on $S^{2}$ (the group of movements of
  the plane in the example, which is \emph{not} commutative). This
  condition means that
  after the above protocol is carried out completely, one always gets the
  original message.
\item\label{cond:big-enough} Conditions on the action. At least the following ones, but more might be needed:
  \begin{itemize}
  \item Given $(s,t)\in S^{2}$ and $g\in G$, for any $s^{\prime}\in S$
    there are $t^{\prime}\in S$ and $g'\in G$ such that $g\cdot (s,t)
    = g^{\prime}\cdot (s^{\prime},t^{\prime})$.
  \item For any $(s,t)in S^{2}$ and $A, B\in G$, there are $(s^{\prime},t^{\prime})$ and $A^{\prime},B^{\prime}\in G$ for which the sequences in the above algorithm are the same:
    \begin{align*}
      [(s,t)\cdot &A, (s,t)\cdot A \cdot B, (s,t) \cdot A \cdot B \cdot
      A^{-1}] = \\ &[(s^{\prime},t^{\prime})\cdot A^{\prime},
      (s^{\prime},t^{\prime})\cdot A^{\prime} \cdot B^{\prime},
      (s^{\prime},t^{\prime})\cdot A^{\prime} \cdot B^{\prime}\cdot
      (A^{\prime})^{-1}].
    \end{align*}
  \end{itemize}
\end{enumerate}
In fact, we do not need exactly an action of $G$ on $S^{2}$.
\begin{definition}
  Let $G$ be a (not necessarily commutative) group acting on a set
  $T$. We say that $t\in T$ is \emph{comm-fixed} if $g\cdot t = t$ for
  any $g\in \text{Comm}(G)$ (the commutator of $G$). A subset
  $S\subset T$ is \emph{comm-fixed} if 
  any $s\in S$ is \emph{comm-fixed}.
\end{definition}
It is clear that a subset $S\subset T$ is comm-fixed if and only if, for any
$s\in S$ and any $g,h\in G$, one has $s=h^{-1}g^{-1}hg\dot s$. From
this, it follows that we do not need exactly an action of a
commutative group on $S^{2}$ but
an action of a (not necessarily commutative) group on a set $X\supset S^{2}$ for which $S^{2}$ is
comm-fixed and which satisfies, at least, condition
(\ref{cond:big-enough}) above.\par
We would like to prove two results; the first one seems relatively easy,
while we have no clue (but are somewhat pessimistic) about the second
one:
\begin{conjecture}\label{con:sufficiency}
  With the above conditions on $X$, $S^{2}$ and $G$, the protocol
  described in section \ref{sec:false-algorithm}
  is information-theoretically safe.
\end{conjecture}
\begin{question}\label{que:existence}
  Do there exist $X,S$ and a group $G$ acting on $X$ for
  which $S^{2}\subset X$ is comm-fixed and such that the stated
  conditions hold?
\end{question}
\textbf{Remark:\/} it is obvious that $S^{2}$ can be changed by any
set of the same cardinal.
\end{document}